\documentstyle[preprint,aps]{revtex}
\tighten
\draft
\begin{document}
\preprint{\today}
\title{Cooling rate effects in amorphous Silica: A Computer
Simulation Study}
\author{Katharina Vollmayr\cite{kvollmayr}, Walter Kob\cite{wkob}
and Kurt Binder}
\address{Institut f\"ur Physik, Johannes Gutenberg-Universit\"at,
Staudinger Weg 7, D-55099 Mainz, Germany}
\maketitle

\begin{abstract}
Using molecular dynamics computer simulations we investigate how in
silica the glass transition and the properties of the resulting glass
depend on the cooling rate with which the sample is cooled.  By
coupling the system to a heat bath with temperature $T_b(t)$, we cool
the system linearly in time, $T(t)=T_i-\gamma t$, where $\gamma$ is the
cooling rate.  We find that the glass transition temperature $T_g$ is
in accordance with a logarithmic dependence on the cooling rate. In
qualitative accordance with experiments, the density shows a local
maximum, which becomes more pronounced with decreasing cooling rate.
The enthalpy, density and the thermal expansion coefficient for the
glass at zero temperature decrease with decreasing $\gamma$. We show
that also microscopic quantities, such as the radial distribution
function, the bond-bond angle distribution function, the coordination
numbers and the distribution function for the size of the rings depend
significantly on $\gamma$. We demonstrate that the cooling rate
dependence of these microscopic quantities is significantly more
pronounced than the one of macroscopic properties. Furthermore we show
that these microscopic quantities, as determined from our simulation,
are in good agreement with the ones measured in real experiments, thus
demonstrating that the used potential is a good model for silica glass.
The vibrational spectrum of the system also shows a significant
dependence on the cooling rate and is in qualitative accordance with
the one found in experiments. Finally we investigate the properties of
the system at finite temperatures in order to understand the
microscopic mechanism for the density anomaly. We show that the anomaly
is related to a densification and subsequent opening of the tetrahedral
network when the temperature is decreased, whereas the distance between
nearest neighbors, i.e. the size of the tetrahedra, does not change
significantly.

\end{abstract}

\narrowtext


\section{Introduction}
\label{sec1}

The last few years have shown that computer simulations are a very
effective tool to gain insight into the structure and dynamics of
supercooled liquids and glasses and that they are therefore a very useful
extension of experimental and analytical investigations of such
systems~\cite{angell81,barrat_rev,kob_rev}. The main reason for the
success of such simulations is based upon two facts: Firstly that they
allow to investigate the structure of such systems in full microscopic
detail and secondly that for most {\it atomic} systems many interesting
dynamical phenomena occur on a time scale that is accessible to such
simulations, i.e. happen between $10^{-12}$ seconds and $10^{-7}$
seconds.  It is this time range on which much of the recent
investigations on the dynamics of supercooled liquids has been focused,
since many of the predictions of the so-called mode-coupling theory, a
theory that attempts to describe the dynamics of supercooled
liquids~\cite{mct}, can be tested well in this time window.

If in a supercooled liquid the temperature is decreased so much that
the relaxation times of the system exceed the time scale of the
experiment or of the computer simulation, the system will fall out of
equilibrium and undergo a glass transition, provided that it doesn't
crystallize. Thus the resulting glass is a nonequilibrium structure and
its properties will in general depend on its history of production such
as, e.g., the rate with which the sample was cooled or compressed.
Such dependencies have indeed been found in experiments and in computer
simulations. For example, it has been demonstrated in experiments
\cite{ritland54,cool_exp,bruning92,bruning94} and in computer
simulations
\cite{fox84,cool_comp,baschnagel93,vollmayr95,lj_cool_let,lj_cool},
that the density or the glass transition temperature depend on the
cooling rate.  In some of these simulations also more microscopic
quantities, such as the radial distribution function or the radius of
gyration of polymers, have been investigated and it was shown that also
these quantities depend on the cooling
rate~\cite{baschnagel93,lj_cool_let,lj_cool}. In particular it was
shown that certain microscopic quantities show a much stronger
dependence on the cooling rate than macroscopic quantities (e.g. in
Ref.~\cite{lj_cool_let,lj_cool}) which shows that it might be
interesting to extend the experiments in this direction also.

An important difference between computer simulations of supercooled
liquids and of glasses should be pointed out. In the former type of
studies one investigates the {\it equilibrium} properties of the
system. Thus a direct comparison between the results from simulations
and experiments is possible.  This is not the case for glasses, which
are {\it non-equilibrium} systems. As mentioned in the previous
paragraph, the temperature at which the system undergoes a glass
transition will depend on the time scale of the experiment. Since the
times scales of the computer simulation are many orders of magnitude
shorter than the ones of a typical laboratory experiment, it follows
that the glass transition temperature on the computer is significantly
higher than the glass transition temperature one observes in the
laboratory (assuming all other things to be equal).  (An exception are
experiments with ion bombardment of glasses in which the cooling rates
become comparable to the ones used in computer
simulations~\cite{klaumunzer92}) Thus if the properties of glasses are
investigated by computer simulations it is necessary to see how these
properties depend on the way the glass was produced before a comparison
with real experiments can be made.  Such a check is of particular
importance if one is interested in the microscopic properties of the
glass since, as we have mentioned above, these quantities usually show
a stronger dependence of the production history than the macroscopic
properties.

The goal of the present paper is twofold. On the one hand we want to
investigate how the cooling rate affects the microscopic properties of
a {\it strong} glass former and compare these dependencies with the
results of a similar simulation we did for a {\it fragile} glass
former~\cite{vollmayr95,lj_cool_let}. Secondly we want to investigate
whether the two-body potential that was recently proposed by van Beest
{\it et al.} (BKS)~\cite{beest_90} for the description of {\it
crystalline} silica, is able to reproduce also structural properties of
{\it amorphous} silica.  Apart from beeing of great importance in
chemistry, geology and industrial applications, silica is also a
prototype of a network forming glass and thus it has been investigated
extensively~\cite{silica_simul,kieffer89,nakado94,no_dens_anomaly,rino_93,badro_95,jin_93,dellavalle_94,mitra82,garofalini84,vollmayr_phd}.
Since the BKS potential contains only two-body terms, it can be
implemented in a simulation much more efficiently than a potential
which contains also three body terms. This in turn allows to make
longer runs and thus to study the equilibrium properties of the system
at lower temperatures or to investigate glasses which have a lower
glass transition temperature and are therefore more realistic.

The rest of the paper is organized as follows. In the next section we
give the details of the used potential as well as of the simulation.
Section~\ref{sec3} contains the results and consists of three parts: In
the first one we study the properties of the system {\it during} the
cooling procedure and therefore the glass transition. In the second
part we investigate how the properties of the glass depend on the
cooling rate with which it was produced, i.e. {\it after} having been
cooled to zero temperature. The third subsection is then devoted to
investigate the system at {\it finite} temperatures in order to relate
the properties of the system in its glass phase to the ones at finite
temperature. In the last section we then summarize and discuss the
results.

\section{Model and Details of the Simulation}
\label{sec2}

One of the most important ingredient for a realistic simulation of
materials is the potential. As already mentioned in the Introduction,
silica is a very important glass former and thus there have been many
investigations in which this system has been studied by means of
computer simulations.  Thus it is not surprising that there are many
different types of potentials in use which seem to be able to give a
more or less realistic description of the real potential. One of the
most successful is the so-called BKS potential, proposed by van Beest
{\it et al} a few years ago~\cite{beest_90}. Is was shown that this
potential is able to give a good description of the various crystalline
phases of silica~\cite{bks_simulations}. It is therefore interesting to
see how well it is able to describe the amorphous phase as well. One of
the appealing features of this potential is that it contains only
two-body terms, thus avoiding the three-body terms that are present in
some other potentials for silica, making the BKS potential very
attractive for computer simulations.

The functional form of the BKS potential is given by a sum of a Coulomb
term, an exponential and a van der Waals term. Thus the potential
between particle $i$ and $j$ is given by

\begin{equation}
\phi(r_{ij})=\frac{q_i q_j e^2}{r_{ij}}+A_{ij}e^{-B_{ij}r_{ij}}-
\frac{C_{ij}}{r_{ij}^6}\quad ,
\label{eq1}
\end{equation}
where $e$ is the charge of an electron and the constants $A_{ij}$,
$B_{ij}$ and $C_{ij}$ are given by $A_{\mbox{\footnotesize SiSi}}=0.0$
eV, $A_{\mbox{\footnotesize SiO}}=18003.7572$ eV,
$A_{\mbox{\footnotesize OO}}=1388.7730$ eV, $B_{\mbox{\footnotesize
SiSi}}=0.0$ \AA$^{-1}$, $B_{\mbox{\footnotesize SiO}}=4.87318$
\AA$^{-1}$, $B_{\mbox{\footnotesize OO}}=2.76000$ \AA$^{-1}$,
$C_{\mbox{\footnotesize SiSi}}=0.0$ eV\AA$^{-6}$,
$C_{\mbox{\footnotesize SiO}}=133.5381$ eV\AA$^{-6}$ and
$C_{\mbox{\footnotesize OO}}=175.0000$ eV\AA$^{-6}$~\cite{beest_90}.
The partial charges $q_{i}$ are $q_{\mbox{\footnotesize Si}}=2.4$ and
$q_{\mbox{\footnotesize O}}=-1.2$ and $e^2$ is given by $1602.19/(4\pi
8.8542)$ eV\AA. The so defined potentials for the Si-O and the O-O
interaction have the unphysical property to diverge to minus infinity
at small distances. However, this is not a severe drawback, since in
order to get to such small distances the particles have to overcome a
barrier which is, e.g., in the case of the Si-O interaction, on the
order of 5000K.  In our simulations we have observed that even at a
temperature of 7000K the particles are relatively unlikely to cross
this barrier, thus indicating that the {\it effective} barrier is
probably even larger than 5000K. In order to prevent that, in the rare
cases in which the particles cross the barrier, the particles fuse
together, we have substituted the potential given by Eq.~(\ref{eq1}) by
a harmonic potential when $r_{ij}$ is smaller than the location of the
barrier, i.e.  for $r_{ij}\leq 1.1936$\AA and $r_{ij}\leq 1.439$\AA in
the case of the Si-O and O-O interaction. Note that for intermediate
and low temperatures this modification does not affect the potential
given by Eq.~(\ref{eq1}) and that in this limit we are thus working
with the usual BKS potential.

The Coulomb interaction was computed by using the Ewald
method~\cite{kieffer89,allen90} with a constant $\alpha/L$ of 6.5,
where $L$ is the size of the cubic box, and by using all $q$-vectors
with $|q|\leq 6\cdot 2\pi /L$.  In order to save computer time the
non-Coulombic contribution to the potential was truncated and shifted
at a distance of 5.5\AA. Note that this truncation is not negligible
since it affects the pressure of the system. We will comment on this
point more when we discuss the temperature dependence of the density.
In order to minimize surface effects periodic boundary conditions were
used. The masses of the Si and O atoms was 28.086u and 15.9994u,
respectively.  The number of particles was 1002, of which 334 were
silica atoms and 668 were oxygen atoms.

Our simulations were done at constant pressure ($p_{ext}=0$), thus
allowing us to compute the temperature dependence of the density and
the specific heat at constant pressure and hence to compare our results
with real experiments. For this we used the algorithm proposed by
Andersen~\cite{andersen80} with the mass of the piston set to $4\cdot
10^{-3}u$ for the equilibration of the system and to $1\cdot 10^{-3}u$
for the production.  The equations of motion were integrated with the
velocity form of the Verlet algorithm. The step size was 1.6 fs which
was sufficiently small to allow us to neglect the drift in the enthalpy
of the system when the thermostat was not active. This thermostat was a
stochastic collision procedure which periodically substituted the
velocities of all the particles with those drawn from a Boltzmann
distribution that corresponded to the temperature of the heat bath. For
the equilibration we coupled the system at every 50 time steps to a
stochastic heat bath and propagated it in the NPT-ensemble at a
temperature of 7000K for about 32,000 time steps. After this time the
configuration and velocities were saved for the subsequent quenching
procedure. Then the equilibration run at T=7000K was continued for
another 40,000 steps and the resulting configuration saved. These
40,000 time steps were long enough to completely decorrelate the system
at this temperature. This process was repeated until we had 20
configurations at $T=7000$K which were completely uncorrelated.

In order to simulate the cooling process we took these configurations
as a starting point of a constant pressure run in which $T_b$, the
temperature of the heat bath, was decreased linearly in time $t$,
i.e.:  $T_b(t)=T_i-\gamma t$. Here $T_i$ is the initial temperature
(=7000K) and $\gamma$ is the cooling rate. The system was coupled to
this heat bath every 150 time steps and between these stochastic
collisions it was propagated in the NPH-ensemble, where H is the
enthalpy. This cooling process was continued until the temperature of
the heat bath was zero, i.e for a time $T_i/\gamma$. The so obtained
configuration was subsequently relaxed with respect to the coordinates
of the particles and the volume of the system to its nearest metastable
state in configuration space. For the sake of efficiency this
relaxation was done with a multi-dimensional conjugate gradient
method~\cite{press92}. An equivalent alternative would have been to
continue the MD simulation at $T_b$=0 for a very long time. The so
obtained final configurations were then analyzed in order to
investigate how the so produced glass depends on the cooling rate.

The cooling rates investigated were: $1.14\cdot 10^{15}$K/s, $5.68\cdot
10^{14}$K/s, $2.84\cdot 10^{14}$K/s, $1.42\cdot 10^{14}$K/s, $7.10\cdot
10^{13}$K/s, $3.55\cdot 10^{13}$K/s, $1.77\cdot 10^{13}$K/s, $8.87\cdot
10^{12}$K/s and $4.44\cdot 10^{12}$K/s.  Although these cooling rates
are of course many orders of magnitude larger than the ones used in the
laboratory it is currently not possible to simulate a
quench of the system with cooling rates that are significantly smaller
than the ones used here, since for the smallest cooling rate the length
of the runs were about $10^6$ MD steps which took about 340 hours of
CPU time on a IBM-RS6000/370.

We also mention that the range of cooling rates investigated here is
about a factor or ten smaller than the one we used in a similar
investigation on a binary Lennard-Jones
mixture~\cite{lj_cool_let,lj_cool}. The reason for this is that for the
Lennard-Jones system the potential is short ranged whereas the long
range potential needed for silica (Eq.~\ref{eq1}) slows down the
computation of the forces by about a factor of 30.

In order to improve the statistics of the results it was necessary to
average for each cooling rate over several independent runs. For most
values of $\gamma$ we averaged over ten independent starting
configurations which were obtained as described above. An exception
were $\gamma=7.10\cdot 10^{13}$K/s and $\gamma=3.55\cdot 10^{13}$K/s
for which we averaged over twenty configurations.

\section{Results}
\label{sec3}

This section consists of three subsections. In the first one we
investigate the properties of the system during the cooling from high
temperatures to zero temperatures, and how the occurring glass
transition depends on the cooling rate. In the second subsection we
study how the properties of the glass at zero temperatures depend on
the cooling rate. In the third subsection we use the information that
we gained in the first two subsections to understand better the
microscopic structure of silica at finite temperatures.

\subsection{Cooling rate dependence of the quench}
\label{sec3_1}

One of the simplest quantities one can study in a cooling process is
the enthalpy $H$ of the system which is given by $H=E_{kin}+E_{pot}+
M\dot{V}^2/2+p_{ext}V$, where $E_{kin}$ and $E_{pot}$ are the kinetic
and potential energy of the system, respectively, and $M$, $V$ and
$p_{ext}$ are the mass of the piston, the volume of the system and the
external pressure. Earlier simulations of glass forming systems have
shown that $H(T_b)$ has a noticeable bend when the temperature is
lowered from high temperatures to low temperatures. It is assumed that
at the temperature at which this bend occurs the system falls out of
equilibrium, because the typical relaxation times of the system exceed
the time scale of the cooling process. Therefore this temperature can
be identified with the glass transition temperature $T_g$.

In Fig.~\ref{fig1} we show the enthalpy of the system as a function of
the temperature of the heat bath for all cooling rates investigated. The
inset shows the whole range of temperature and we see that the curves
show the mentioned bend at a temperature around 3500K. This is thus the
temperature range in which the system falls out of equilibrium for the
cooling rates investigated. This temperature range is shown enlarged in
the main figure. We now see that there is a clear dependence of
$H(T_b)$ on the cooling rate in that the curves corresponding to the
large cooling rates are higher than the ones for the small cooling
rates. At high temperatures the curves for intermediate and small
cooling rates fall on top of each other to within the noise of the
data, which means that for these temperatures and cooling rates the
system has not yet fallen out of equilibrium. Only at lower temperatures do
the curves for the intermediate values of the cooling rate split off
from this equilibrium (liquidus) curve and thus is the system starting
to undergo a glass transition and we see that the temperature at which
this happens decreases with decreasing cooling rate. Also note that for
the largest cooling rates this splitting off happens at the starting
temperature, thus indicating that for such large cooling rates the
system falls out of equilibrium immediately.

In order to determine the cooling rate dependence of the temperature at
which the system undergoes its glass transition we use the concept of
the ``fictive temperature'' as introduced by Tool {\it et
al}~\cite{tool31}. This concept makes use of the observation that at
high temperatures the curves for not too large cooling rates fall onto
a master curve and that at low temperatures the curves have the same
form, i.e. can be collapsed onto a master curve by shifting them
vertically. The intersection of the extrapolation of these two master
curves gives then an estimate for the glass transition temperature
$T_g$. We therefore fitted the curves for $\gamma \leq 3.55\cdot
10^{13}$K/s in the temperature range 5000K $\leq T \leq$ 6750K with a
straight line and did the same with the curves for $\gamma \leq 1.42
\cdot 10^{14}$K/s in the low temperature range 0 $\leq T \leq$ 1250K.
Note that the determination of the glass temperature via the mentioned
procedure is only reasonable if the high temperature part of the curve
actually falls onto the liquidus curve. Since this is not the case for
the three fastest cooling rates we have not determined $T_g$ for these
cooling rates.

In Fig.~\ref{fig2} we show the so determined glass transition
temperature as a function of the cooling rate. We see that a variation
of $\gamma$ by about 1.5 decades gives rise to a variation of $T_g$ of
about 350K. Also included in the figure is a fit to the data with the
functional form

\begin{equation}
T_g(\gamma)=T_0-\frac{B}{\ln (\gamma A)}
\label{eq2}
\end{equation}
(solid line), which is obtained by assuming a Vogel-Fulcher dependence
of the relaxation time $\tau$ of the system on the temperature, i.e.
$\tau(T)=A\exp(B/(T-T_0))$, and arguing that the system falls out of
equilibrium at that temperature at which the relaxation time is on the
order of the time scale of the cooling process, i.e.  $\tau(T_g)=
\gamma^{-1}$~\cite{angell81,angell83}. We see that this type of fit
describes the data very well, as it is the case in real
experiments~\cite{bruning92}. For the parameters $A$ and $B$ we find
$1.8\cdot 10^{-16}$s/K and 2625K, respectively. The Vogel temperature
$T_0$, i.e. the glass temperature that would be observed upon an
infinitesimal cooling rate, is 2525K, which is significantly higher
than the experimental value of 1446K~\cite{angell88}.  (Here we assume
that for the cooling rates used in the laboratory the dependence of
$T_g$ on the cooling rate is sufficiently small, so that we can use the
results of experiments at a finite cooling rate as a good approximation
for $T_0$.) Thus we come to the conclusion that either the
extrapolation of the cooling rate dependence as given in
Eq.~(\ref{eq2}) is not correct, because we are not yet in the range of
cooling rates where Eq.~(\ref{eq2}) holds, or that the silica model
studied here does not reproduce well the glass transition temperature.

We also mention that a further source of uncertainty in the
determination of $T_0$ are finite size effects in the simulation. It
has been shown that for long ranged potentials the structure shows
noticeable finite size effects~\cite{nakado94}, thus it may well be
that also $T_g$ shows a dependence on the system size.  Such
dependencies have indeed been observed in experiments with relatively
simple liquids~\cite{glasses_pores} and also a recent computer
simulation of silica has shown that the relaxation behavior of such a
system is severely affected by finite size effects~\cite{horbach96}.

In a similar study on cooling rate effects in a Lennard-Jones
glass~\cite{lj_cool} we have found that the dependence of $T_g$ on the
cooling rate is also fitted well by the function
$T_g(\gamma)=T_c+(A\gamma)^{1/\delta}$, which follows from the
assumption that the temperature dependence of the relaxation time is
given by $\tau(T)=A(T-T_c)^{-\delta}$, a functional form that is
suggested by the so-called mode-coupling theory of the glass
transition~\cite{mct}. We therefore tried to fit our data for the
$\gamma$ dependence of $T_g$ also in the present case with this
functional form and found that it is also able to describe the data
well (with $T_c=2778$K and $\delta=2.52$) (dashed line in
Fig.~\ref{fig2}). Thus if the two functional forms are merely seen as
fitting functions they can be considered as equally good.  However,
since it is commonly assumed that the temperature dependence of the
relaxation time of real silica is much better described by a
Vogel-Fulcher law than by a power-law, it is reasonable to give in the
present case the preference to the former functional form. (On the
other hand, R\"ossler and Sokolov have demonstrated recently that also
for silica the viscosity at temperatures a bit above $T_g$ shows a
non-Arrhenius behavior, which would support the idea of a power-law
fit~\cite{rossler96}.)

As stated above, we observe a change of $T_g$ of about 300K when the
cooling rate is varied by 1.5 decades.  Such a change in $T_g$ is
significantly larger than the values measured in real experiments in
which this quantity was determined for various
materials~\cite{ritland54,bruning92,bruning94}. It is found that a
variation of the cooling rate by one decade gives rise to a change of
$T_g$ on the order of 10K thus much less than the 300K determined
here.  The reason for this discrepancy is probably the huge difference
between the  cooling rates used in the simulation and the one used in
the laboratory. If we use the parameters from our fit to $T_g$ and
extrapolate this $T_g(\gamma)$ dependence to a laboratory cooling rate
of 0.1K/s, we find that the predicted change of $T_g$ is only about 5K
per decade of cooling rate, which is in good agreement with the typical
values found in experiments~\cite{ritland54,bruning92,bruning94}.

By differentiating the enthalpy with respect to the temperature $T_b$,
we obtain $c_p$, the specific heat at constant pressure. Since the
original data was a bit too noisy to allow for a direct
differentiation, we parametrized $H(T_b)$ with a spline under
tension~\cite{reinsch67} and differentiated this spline.
Figure~\ref{fig3} shows the resulting specific heat for all cooling
rates investigated. To facilitate the comparison with experimental
values we have chosen the units of $c_p$ to be J/gK. From this figure
we see that at high temperatures the fastest cooling rates show a
strong increase with decreasing temperature. The reason for this is
that for these fast cooling rates the system falls out of equilibrium
already at the start of the quench (see Fig.~\ref{fig1}) thus giving
rise to this increase of $c_p$. After this increase $c_p$ attains a
maximum of about 1.85 J/gK at a temperature around 4200K and then drops
again to a value of 1.25J/gK for $T_b=0$.

The real equilibrium curve of $c_p$ at high temperatures is given by
the curves for slow cooling rates. We see that $c_p$ increases slowly
from a value around 1.80 J/gK to a value around 1.95 J/gK when the
temperature is decreased from 7000K to 4300K. At this latter
temperature the specific heat starts to drop quickly, indicating that
the system undergoes the glass transition, and attains a value around
1.25 J/gK at $T_b=0$K. This value is close to the classical
Dulong-Petit value of 1.236 J/gK expected for a harmonic solid. We
notice that in the temperature range where the glass transition takes
place the temperature dependence of $c_p$ is independent of the cooling
rate to within the accuracy of our data, if the cooling rate is not too
large.  This is in contrast to our findings for the previously
investigated Lennard-Jones system~\cite{lj_cool}, for which we found
that the drop in $c_p$ at the glass transition becomes steeper with
decreasing cooling rate. Since we have seen a clear cooling rate
dependence of the temperature dependence of the enthalpy, see
Fig.~\ref{fig1}, it can be concluded that its derivative, i.e. $c_p$
should show a cooling rate dependence also and that thus the reason for
our failure to detect one must be given by the statistical inaccuracy
of our data.

Since neither the low nor high temperature dependence of $c_p$ show a
strong dependence on the cooling rate, if $\gamma$ is not too large, we
can compare the values of $c_p$ above and below the glass transition
temperature with their experimental counterparts.  Br\"uckner reports
that around $1500K$ the value of $c_p$ for
amorphous silica is about 1.23 J/gK~\cite{bruckner_70}, which compares
well with the one found in this simulation, i.e. 1.25 J/gK. At a
temperature of 2000K Br\"uckner gives the value 1.50 J/gK, which is
significantly less than the one found in our simulation ($c_p\approx
1.8$ J/gK at $T=7000$K). Thus we see that the here used BKS potential does
not give an accurate description of the magnitude of the jump in the
specific heat.  One possible reason for the observed discrepancy might,
however, not be the inadequacy of the potential, but the fact that the
Debye temperature of silica is relatively high
(1200K~\cite{bruckner_70}), thus showing that quantum effects might be
important even at the temperatures we are considering.

We now turn our attention to a further important macroscopic quantity,
the density $\rho$. In Fig.~\ref{fig4} we show $\rho$ as a function of
the bath temperature $T_b$ for all cooling rates investigated. As in
the case of the enthalpy we find that at high temperatures the curves
for all but the three fastest cooling rates fall onto a master curve, the
equilibrium curve. From the
curves corresponding to small cooling rates we recognize that this
equilibrium curve shows a maximum at around 4800K. Thus we find that,
in accordance with experiments~\cite{bruckner_70,mazurin83,angell76},
this model shows an anomaly in the density. The experimental value for
the temperature of the maximum in $\rho$ is 1500K, thus significantly
lower than the temperature at which we observe the anomaly. Since we
see that, within the accuracy of our data, the temperature at which
this anomaly occurs is independent of the cooling rate we conclude that
for the BKS potential this anomaly is indeed at a temperature which is
too high, even if one would cool the system with a significantly
smaller cooling rate.  It has to be mentioned, however, that for
different potentials this anomaly occurs at even higher temperatures or
is not present at all \cite{no_dens_anomaly} thus showing that with
respect to this feature the BKS potential is superior to other
potentials.

For intermediate and small values of $\gamma$ the value of $\rho$
decreases after having passed through the maximum. At even lower
temperatures the curves then start to increase again. The temperature
at which this increasing trend starts, decrease with decreasing cooling
rate thus showing that the curves follow the low temperature side of the
hump the longer the smaller the cooling rate is. At even lower
temperatures the curves become, within the accuracy of our data,
straight lines with negative slope.

From Fig.~\ref{fig4} we also recognize that, in the temperature range
considered, the relative change in density is relatively small (less
than 10\%) which is in accordance with the experimental finding that
the thermal expansion coefficient of silica is
small~\cite{mazurin83,rubber_book}. We also note that at low
temperatures the density is around 2.3 g/cm$^3$, which compares well
with experiments~\cite{mozzi_69}.  It is interesting that a simulation
with the {\it original} BKS potential, i.e. without the cutoff at
5.5\AA (see Sec.~\ref{sec2}), gave at 5000K a density of 2.6g/cm$^3$,
thus about 10\% higher than the one found in this
work~\cite{barrat_priv}. This shows how sensitively quantities like the
pressure depend on the details of the potential at large distances.
Since the introduction of the cutoff moves the value of the density
closer to the experimental one, we thus find that this cutoff gives
rise to a more realistic description of amorphous silica.

From the temperature dependence of the density we can extract
\begin{equation}
\alpha_p=\frac{1}{V}\left.\frac{\partial V}{\partial T}\right|_p =
-\frac{1}{\rho} \left.\frac{\partial \rho}{\partial T}\right|_p 
\quad ,
\end{equation}
the thermal expansion coefficient at constant pressure. The temperature
dependence of $\alpha_p$ is shown in Fig.~\ref{fig5} for all cooling
rates investigated. As in the case of the specific heat, we find that
for large cooling rates $\alpha_p$ shows an unphysical behavior at high
temperatures, in that it increases with decreasing temperature. The
reason for this behavior is again that for these cooling rates the
system falls out of equilibrium almost immediately. For the
intermediate and small values of the cooling rate we see that
$\alpha_p$ decreases with decreasing temperature, becomes negative and
then starts to increase again before it goes to a finite positive value
at small temperatures. The reason for this non-monotonous behavior is
the presence of the anomaly in the density, which gives rise to the
negative expansion coefficient. From the figure we also recognize that
the minimum in $\alpha_p$ becomes more pronounced the smaller the
cooling rate is, in accordance with our earlier observation, commented
in the discussion of Fig.~\ref{fig4}, that the smaller the cooling rate
is, the longer the curves for $\rho(T_b)$ follow the low temperature side
of the density anomaly to smaller values of $\rho$.

From this figure we also recognize that the expansion coefficient
at $T=0$K seems to decrease with decreasing cooling rate. In order to
investigate this behavior more thoroughly we determined $\alpha_p$ from
the low temperature dependence of the density, i.e. from the slope of
the straight line of $\rho(T)$ at low temperatures (see
Fig.~\ref{fig4}). The resulting cooling rate dependence of
$\alpha_p(T=$0 K) is show in Fig.~\ref{fig6}. We recognize that this
quantity shows a decreasing trend with decreasing cooling rate and
that, within the accuracy of our data, it is not possible to say what
the asymptotic value for very small cooling rates is. However, the
experimental value of $\alpha_p$, $5.5 \cdot 10^{-7}$
K/s~\cite{mazurin83,rubber_book} is certainly compatible with a
extrapolation of our data to $\gamma=0$.

It is also interesting to compare this result with the one found in a
similar investigation of a binary Lennard-Jones system~\cite{lj_cool},
where {\it no} significant dependence of $\alpha_p$ at $T=0$K on the
cooling rate was found. Since a nonzero $\alpha_p$ is the result of the
anharmonicity of the local potential we thus come to the conclusion
that in silica these anharmonic effects are cooling rate dependent,
whereas they are not for the Lennard-Jones system, i.e. for a prototype
of a simple liquid.

\subsection{Cooling rate dependence of the properties of the glass}
\label{sec3_2}

In the previous subsection we investigated how the cooling rate affects
macroscopic quantities like the enthalpy or the density at {\it finite}
temperatures. The goal of the present subsection is to see how the
cooling rate affects various macroscopic and microscopic quantities of
the glass at $T=0$K, i.e. of the final product of the quench and the
subsequent relaxation of the system as described above.

The first quantity we investigate is the value of the enthalpy of the
glass at $T=0$K. In Fig.~\ref{fig1} we have seen that at finite
temperatures the curves of the enthalpy $H(T_b)$ follow the equilibrium
curve as long as the relaxation time of the system is smaller than the
time scale of the cooling process, i.e. $\gamma^{-1}$. If the two time
scales become comparable the system undergoes a glass transition and
the curve for $H(T_b)$ remain {\it above} the equilibrium curve.
Therefore we expect that the final value of the cooling process
decreases with decreasing $\gamma$. That this is indeed the case is
shown in Fig.~\ref{fig7} where we show $H_f$, the value of the
enthalpy after the quench, for all cooling rates investigated.

From a formal point of view the cooling process can also be seen as an
optimization problem in which the system tries to minimize the
enthalpy.  It will manage to do this the better the more time it is
given to search for this minimum. Thus one might ask what the value of
the enthalpy (the cost function) is when the system is given a certain
amount of time, characterized here by the cooling rate, to minimize $H$.
Such types of questions have been addressed already in other types of
complex optimization problems and also for other types of glass
formers~\cite{kob90,grest_huse,new_optimization}. From theoretical
arguments one can expect the cost function to show
either a logarithmic or a power-law dependence on the cooling
rate~\cite{grest_huse}, i.e.  
\begin{equation}
H_f(\gamma)=H_f^0+a_1(-\log \gamma)^{a_2} 
\label{eq3} 
\end{equation} 
or
\begin{equation} 
H_f(\gamma)=H_f^0+b_1\gamma^{b_2} \qquad,
\label{eq4}
\end{equation} 
where $H_f^0$, the $a_i$ and $b_i$ are fit parameters.
We therefore fitted our data for $H_f$ with the two functional forms
and the result of these fits is included in Fig.~\ref{fig7} as well. We
recognize that in the cooling range investigated both functional forms
fit the data equally well. The values we obtain for $H_f^0$ is -19.1213eV
for the logarithmic dependence and -19.1252eV for the power-law
dependence. With the accuracy of our data we are not able to
decide which functional form, if any, is appropriate to describe our
data. This is the same conclusion we came to in case of the earlier
mentioned investigation of a Lennard-Jones system~\cite{lj_cool}.

Next we turn our attention to the cooling rate dependence of the
density. In Fig.~\ref{fig8} we show the density of the glass after the
quench versus the cooling rate. The densities we found for the glass
are between 2.27g/cm$^3$ and 2.38g/cm$^3$ which compares well with the
experimental values of 2.2g/cm$^3$\cite{mazurin83,mozzi_69}. We see
that, contrary to most real glasses or Lennard-Jones
systems~\cite{ritland54,bruning94,lj_cool}, $\rho_f$ {\it decreases}
with decreasing cooling rate, a behavior that can be understood by
remembering our observation (see Fig.~\ref{fig4}) that for small
cooling rates the curves for the density follow the equilibrium curve
also when the latter is decreasing on the low temperature side of the
density anomaly. It should be noted, however, that this observed
decrease of the final density with decreasing cooling rate cannot be
the correct asymptotic behavior for very small cooling rate. The reason
for this is that we know that at low temperatures the thermal expansion
coefficient of silica is positive (as can be seen in experiments or
from the fact that the density decreases with increasing temperature
(see Fig.~\ref{fig4})). Thus we expect that the equilibrium curve for
the density will, after having shown a decreasing behavior for
temperatures just below the density anomaly, bend upward again. If a
quench is made with a very small, but finite, value of $\gamma$ the
corresponding curve for the density will fall out of equilibrium in
that temperature range where the equilibrium curve will already show
the {\it increasing} behavior (with decreasing temperature). Therefore
the final density of the glass as produced with such a small cooling
rate will be {\it lower} than the one which would be obtained with an
infinitesimal small cooling rate, i.e. at very small cooling rate the
curve $\rho_f(\gamma)$ will increase with decreasing cooling rate. Thus
we conclude that the cooling rate dependence of $\rho_f$ as seen in
Fig.~\ref{fig8} is not yet the asymptotic one. Hence it does not make
sense to use one of the formulae given in Eqs.~(\ref{eq3}) and
(\ref{eq4}) to extrapolate $\rho_f(\gamma)$ to very small cooling
rates.

After having presented our results on the cooling rate dependence of
the macroscopic properties of the glass we now turn our attention to the
microscopic properties of the system in order to gain some understanding
about how the macroscopic behavior is related to the microscopic one.

The first quantity we investigate is the radial distribution function
$g_{\alpha\beta}(r)$ between species $\alpha$ and $\beta$ ($\alpha,
\beta \in \{\mbox{Si,O}\})$~\cite{hansen_mcdo}. This function allows to
see how the structure of the glass changes on the various length scales
when the cooling rate is changed. In Fig.~\ref{fig9}a we show
$g_{\mbox{\footnotesize SiSi}}(r)$ for the largest and smallest cooling
rate investigated (main figure), as well as an enlargement of the
region of the second nearest neighbor peak for a few selected cooling
rates (inset). From the main figure we recognize that with decreasing
cooling rate the order at short and intermediate distances (i.e. $r\leq
8$\AA) increases, in that the peaks and minima become more pronounced.
In particular we see that the height of the first nearest neighbor peak
changes by about 20\%. The amount of this change is significantly
larger than any change we observed for macroscopic properties, thus showing
that the microscopic properties can show a much stronger dependence on
the cooling rate than the macroscopic properties do.

In Fig.~\ref{fig9}b we show the radial distribution function for the
Si-O and the O-O pairs for the largest and smallest cooling rate
investigated. Also in this case we notice a significant cooling rate
dependence for distances $r\leq 8$\AA, i.e. the short and medium
range order are affected significantly by the cooling rate in that
the order increases with decreasing cooling rate. 

From the curves presented in Fig.~\ref{fig9} we see that, although
the {\it height} of the various peaks shows a significant dependence on
the cooling rate, the {\it location} of the peaks is affected much less
by a variation of $\gamma$. Thus it reasonable to compare the location
of these peaks with the ones as determined in experiments.  In
Tab.~\ref{tab1} we give the location of the nearest and second nearest
neighbor peak as well as the corresponding experimental values (also
included in Fig.~\ref{fig9} as vertical dotted lines). The location of
these peaks were determined from the data for the slowest cooling
rate.  We see that, although the accordance between experiment and the
results of our simulation is not perfect, the BKS potential does quite
well to reproduce the short and medium range structure of the glass and
can therefore, from this point of view, be considered as a good model
also for amorphous silica.

Having investigated the cooling rate dependence of the radial
distribution function we now move on to study how the structure factor
$S(q)$ depends on $\gamma$. Although from a mathematical point of view
the radial distribution function and the structure factor contain the
same information, the importance of the latter for scattering
experiments makes it worthwhile to investigate its cooling rate
dependence as well. In Fig.~\ref{fig10} we thus show the three partial
structure factors.  In order not to crowd the figures too much we show
only the curves for the fastest and slowest cooling rates. We recognize
from these figures that the $S(q)$ show a significant dependence on the
cooling rate for small and intermediate values of $q$, in that the
height of the main peak as well as the so-called first sharp
diffraction peak (FSDP), i.e. the peak to the left of the main peak,
depend on $\gamma$. This FSDP has recently been the focus of
significant interest, since it characterizes the stucture of the glass
on intermediate length scales and its microscopic origin is still a
matter of debate~\cite{fsdp}.  In the case of the Si-Si and the Si-O
correlation the corresponding structure factors show only a weak
dependence on $\gamma$ for $q$-values larger than the location of the
main peak. For the case of the O-O correlation, however, even for large
values of $q$ a noticeable dependence of $S(q)$ on $\gamma$ is
observed, indicating that the short range order of O-O pairs changes
significantly. The main change in the structure of $S(q)$ occurs,
however, in all three correlation function for values of $q$ close to
the FSDP (see insets).  We see that the cooling rate affects this peak
in two ways in that its height as well its position is changed. Since
this peak reflects the medium range order of the system, we thus come
to the conclusion that the structure of the glass on these length
scales is significantly affected by the cooling rate.

From the knowledge of $r_{\mbox{\footnotesize min}}$, the location of
the first minimum in the radial distribution function, we can compute
the (partial) coordination number $z$ of particle $i$, which we define
as the number of other particles $j$ with $|\mbox{\boldmath
$r$}_j-\mbox{\boldmath $r$}_i| <r_{\mbox{\footnotesize min}}$. We have
found that $r_{\mbox{\footnotesize min}}$ is essentially independent of
the cooling rate and therefore we will use in the following always the
same values, i.e.  $r_{\mbox{\footnotesize min}}^{\mbox{\footnotesize
SiSi}}=3.42$\AA, $r_{\mbox{\footnotesize min}}^{\mbox{\footnotesize
SiO}}=2.20$\AA  and $r_{\mbox{\footnotesize min}}^{\mbox{\footnotesize
OO}}=3.00$\AA. In Fig.~\ref{fig11} we show $P_{\alpha\beta}^{z=n}$, the
probability that a particle of type $\alpha$ has $n$ nearest neighbors
of type $\beta$, versus the cooling rate $\gamma$.

First we study the nearest neighbor pairs, i.e. the Si-O and the O-Si
pairs (Figs.~\ref{fig11}a and \ref{fig11}b). We see that the vast
majority of the silicon atoms are surrounded by 4 oxygen atoms, which
can be understood by taking into account that at zero pressure silica
forms a network of corner-sharing tetrahedra, each of which has a
silicon atom in its center and four oxygen atoms at its corners. The
number of silica atoms that are not fourfold coordinated is about 5\%
for the fastest cooling rate and diminishes quickly to less than 0.5\%
when the cooling rate is decreased. This shows that the local order of
the network, i.e. the frequency of tetrahedra, increases fast with
decreasing cooling rate. The silicon atoms that are not fourfold
coordinated are surrounded in most cases by five oxygen atoms and only
a very small fraction is surrounded by three oxygen atoms.
Figure~\ref{fig11}b shows that most of the oxygen atoms are surrounded
by two silicon atoms, also this observation in accordance with the
above mentioned network structure of corner sharing tetrahedra. The
number of oxygen atoms that are not twofold coordinated is for all
cooling rates less than 3\% and decreases quickly to less than 0.5\%
with decreasing $\gamma$. Thus we find that for slow cooling rates the
BKS potential automatically gives the ``rules'' commonly postulated for
ideal amorphous silica, namely that this system is a ``continous random
network''~\cite{zallen83}.

The just studied Si-O and O-Si coordination numbers are characteristic
for the structure of the network on the {\it shortest} length scale.
The coordination numbers for the Si-Si and the O-O pairs, however, are
sensitive on a length scale of the structure that is a bit larger. In
Figs.~\ref{fig11}c and~\ref{fig11}d we show the cooling rate dependence
of these coordination numbers. We see that most silicon atoms are
surrounded by four other silicon atoms, although at the fastest cooling
rate about 18\% of them have a different Si-Si coordination number.
This number shows that most tetrahedra are surrounded by four other
tetrahedra, each of which has a silicon atom in its center. (See below
for a further discussion of this point.) The curves for the O-O
coordination numbers show that the most likely configuration is that an
oxygen atoms has six other oxygen atoms within a distance of
$r_{\mbox{\footnotesize  min}}^ {\mbox{\footnotesize OO}}$ and that
this probability increases significantly with decreasing cooling rate.
These six oxygen atoms are the ones that sit in the corners of the two
tetrahedra which are connected by the first oxygen atom (see
Fig.~\ref{fig12}). Thus we come also in this case to the conclusion
that the order within the network increases with decreasing cooling
rate.

Since we have just seen that the most frequent coordination number for
the Si-Si and the O-O pairs can be rationalized by assuming that the
network is composed of corner-sharing tetrahedra, we now investigate
whether this argument is valid only on a qualitative basis or whether
it is correct even on a quantitative basis. Thus the question is
whether the cooling rate dependence of the various coordination numbers
for the Si-Si and the O-O pairs can be computed from the knowledge of
the cooling rate dependence of the coordination numbers of the Si-O and
the O-Si pairs. In order to decide this we assumed that the
coordination numbers for the Si-O are statistically independent from
the ones of the O-Si pairs. If we postulate that a silicon atom
that is a nearest neighbor of an oxygen atom will have a distance less
than $r_{\mbox{\footnotesize min}}^{\mbox{\footnotesize SiSi}}$ from
all other silicon atom that are also nearest neighbors of this oxygen
atom, it is relatively simple to compute the probability how many
silicon atoms have a distance less than $r_{\mbox{\footnotesize min}}^
{\mbox{\footnotesize SiSi}}$ from any given silicon atom. Using similar
postulates for the other combinations of particles one can, e.g., show
that within this ansatz the quantity $P_{\mbox{\footnotesize
SiSi}}^{z=4}$ is given by

\begin{eqnarray}
P_{\mbox{\footnotesize SiSi}}^{z=4} & = & P_{\mbox{\footnotesize SiO}}^{z=3} 
\left[
3 P_{\mbox{\footnotesize OSi}}^{z=1} 
(P_{\mbox{\footnotesize OSi}}^{z=3})^2 + 
3 (P_{\mbox{\footnotesize OSi}}^{z=2})^2 
P_{\mbox{\footnotesize OSi}}^{z=3}
\right]  \nonumber\\
& &
+ P_{\mbox{\footnotesize SiO}}^{z=4} \left[
6 (P_{\mbox{\footnotesize OSi}}^{z=1})^2 
(P_{\mbox{\footnotesize OSi}}^{z=3})^2 +
4 (P_{\mbox{\footnotesize OSi}}^{z=2})^4 +
12 P_{\mbox{\footnotesize OSi}}^{z=1} 
(P_{\mbox{\footnotesize OSi}}^{z=2})^2
P_{\mbox{\footnotesize OSi}}^{z=3}
\right] \nonumber\\
& &
+ P_{\mbox{\footnotesize SiO}}^{z=5} \left[
5 P_{\mbox{\footnotesize OSi}}^{z=1} 
(P_{\mbox{\footnotesize OSi}}^{z=2})^4 +
30 (P_{\mbox{\footnotesize OSi}}^{z=1})^2
(P_{\mbox{\footnotesize OSi}}^{z=2})^2 
P_{\mbox{\footnotesize OSi}}^{z=3}+
10 (P_{\mbox{\footnotesize OSi}}^{z=1})^3
(P_{\mbox{\footnotesize OSi}}^{z=3})^2
\right] \quad .
\label{eq5}
\end{eqnarray}
Similar expressions hold for the other values of the coordination
numbers shown in Fig.~\ref{fig11}c and~\ref{fig11}d. Equipped with
these functions we now can compare the prediction of this factorization
approximation with the measured values for the coordination numbers. In
Fig.~\ref{fig13} we show the difference between the actual value of the
coordination numbers and the predicted ones, i.e.
$P_{sim}^z-P_{appr}^z$. We recognize from Fig.~\ref{fig13}a, that for
the Si-Si pairs this factorization approximation is very good in that
the difference between the actual values and the predicted one is less
than 1.5\% for fast cooling rates and is essentially zero, to within
the statistical accuracy of our data, for small cooling rates. Thus we
find that this factorization approximation works very well for the
Si-Si pairs.

This agreement between the real data and the factorization
approximation is not as good for the case of the O-O pairs
(Fig.~\ref{fig13}b). We see that the discrepancy can be as large as
25\% for the fastest cooling rate but that it diminishes, however, to
less than 13\% for the slowest cooling rate. The reason that the
factorization approximation does not work as well in this case as it
did in the case of the Si-Si pairs is likely to be the fact that two
corner-sharing tetrahedra are tilted towards each other, i.e. the angle
between the silicon atom \#1, the bridging oxygen atom \#3, and the
silicon atom \#2 (see Fig.~\ref{fig12}) is significantly less than
$180^{\circ}$.  Therefore the oxygen atom \#2 is also quite close to
the oxygen atom \#1, although the former is, from a topological point
of view, quite far away from the latter. Therefore it is not unlikely
that the oxygen atom \#1 will have more than just six nearest neighbor
oxygen atoms. This is confirmed by the curve $P_{\mbox{\footnotesize
OO}}^{z=7}$ (see Fig.~\ref{fig11}d), which shows that for fast cooling
rates more than 20\% of the oxygen atoms have more than six other
oxygen atoms as nearest neighbors, and that this figure does not drop
below 10\% even in the case of the slowest cooling rate. That the
tetrahedra actually have the local arrangement suggested above can be
inferred from the bond-bond-angles between neighboring atoms and
therefore we will investigate this quantity next.

We have seen in Fig.~\ref{fig9} that the nearest neighbor distance
between silicon and oxygen atoms is essentially independent of the
cooling rate. Thus we conclude that the tetrahedra do not change their
size significantly when the cooling rate is varied. However, since we
have found that the density of the system depends on the cooling rate,
it must therefore be the case that it is the {\it relative arrangement}
of neighboring tetrahedra which changes with $\gamma$.  One possibility
to characterize this relative arrangement is to consider the various
bond-bond-angles between the different atoms. In Fig.~\ref{fig14} we
show the cooling rate dependence of the distribution function for some
selected angles for various cooling rates. Figure~\ref{fig14}a shows
this distribution function for the tetrahedral angle O-Si-O for all
cooling rates investigated. For a perfect tetrahedra this angle is
109.47$^{\circ}$. We see that $P_{\mbox{\footnotesize OSiO}}$ has
indeed a maximum close to this ideal angle. The location of this
maximum is not quite the one of the ideal tetrahedron but with
decreasing cooling rate it approaches this value.  A decreasing cooling
rate also leads to an increase of the height of the peak as well as a
decrease of its width.  Thus we find that the structure of the local
tetrahedra approaches indeed the one of an ideal tetrahedron when the
cooling rate is decreased. The position of this peak is also in good
agreement with the position found in experiments as can be seen from
Tab.~\ref{tab2}.

In Fig.~\ref{fig14}b we show the distribution function for the angle
between three neighboring oxygen atoms, $P_{\mbox{\footnotesize OOO}}$,
for the fastest and slowest cooling rate investigated (dashed and solid
curve, respectively). We see that $P_{\mbox{\footnotesize OOO}}$ has
two peaks. The first one is relatively sharp and has its maximum around
60$^{\circ}$. It corresponds to the angle that is formed by three
oxygen atoms of the same tetrahedron (e.g.  O\#1-O\#3-O\#4 in
Fig.~\ref{fig12}). With decreasing cooling rate this peak becomes
significantly higher and narrower, indicating that the tetrahedra
become more regular. The second peak is much broader than the first one
and is located at around 135 degrees. Its position changes from around
128$^{\circ}$ for fast cooling rates to around 137$^{\circ}$ for slow
cooling rates. This peak corresponds to angles that are formed by an
oxygen on one tetrahedron, a bridging oxygen and a third oxygen on the
second tetrahedron (e.g. O\#1-O\#3-O\#2 in Fig.~\ref{fig12}). (Note
that this angle is not only sensitive to the relative position of the
two tetrahedra, i.e. the angle Si\#1-O\#3-Si\#2, but also to their
relative orientation. If the upper tetrahedron in Fig.~\ref{fig12} is
rotated around the axis given by O\#3-Si\#2 the mentioned angle between
the three O atoms will change also. This is probably the reason why the
second peak is so broad.) The fact that this angle widens up with
decreasing cooling rate shows that the two neighboring tetrahedra move
away from each other, thus making the structure less dense. Thus this
mechanism is presumably the reason for the decrease in density after
the density anomaly (see Fig.~\ref{fig4}). We will investigate this
point more in Sec.~\ref{sec3_3}.

The picture of an opening network with decreasing cooling rate is also
corroborated by the distribution of the Si-O-Si angle which is included
in Fig.~\ref{fig14}b also. We see that for fast cooling rates this
distribution shows a large peak at 141$^{\circ}$ whose position moves
to 152$^{\circ}$ for the smallest cooling rate, thus indicating that
the network is opening up. From Tab.~\ref{tab2} we recognize that 
at the smallest cooling rate the location and the width of the peak
are in fair agreement with the experimental values.

The angles O-Si-O, O-O-O and Si-O-Si measure the angles between
particles that are located on one or two tetrahedra. The fourth angle
we consider, Si-Si-Si, is, however, defined by three particles that are
in the center of {\it three} tetrahedra. Thus this angle is sensitive
to the structure of the network on a bit larger scale. In
Fig.~\ref{fig14}b be show the distribution function for this angle as
well.  We see that this distribution shows a small peak at about
60$^{\circ}$ and a large peak between 90$^{\circ}$ and 120$^{\circ}$
which seems to be composed of at least two, possibly even three, peaks.
The small peak at 60$^{\circ}$ was also observed by Rino {\it et
al}~\cite{rino_93} and in that paper it was shown that such an angle
occurs when the rings in the network (defined below) have length four.
Since such short rings occur relatively seldom, see below, also the
corresponding peak is small. We also see that the height of this peak
decreases significantly with decreasing cooling rate, which is in
accordance with the observation discussed below that the number of
rings of length four decreases with decreasing cooling rate. Also the
main peak shows a noticeable dependence on $\gamma$ in that its height
increases and its width decreases with decreasing $\gamma$. Because, as
mentioned above, this angle characterizes the structure of the network
on intermediate length scale, it is difficult to draw conclusions about
the nature of this structure from the cooling rate dependence of this
distribution and thus we do not attempt to do it at this point.

A different way to characterize the structure of a network on the
intermediate length scale is to consider the distribution of the
frequency of rings of a given size. A ring is defined as follows:
Starting from a Si-atom one chooses two different O-atoms that are
nearest neighbors. Pick one of these. In general this O-atoms will also
be nearest neighbor of a second Si-atom. From this new Si-atom one then
picks a new nearest neighbor O-atom etc. This process is continued
until one returns to the O-atom which is the second one of the nearest
neighbor O-atoms of the first Si-atom. In this way one has constructed
a closed loop of Si-O segments. The shortest one of these loops is
called the ring associated with the original Si-atom and the two
nearest neighbor O-atoms. The number of Si-O segments in this loop is
called the size of this ring. Both, the distribution with which the
so defined rings occur, as well as the distances and bond angles
present in these rings were studied extensively in the paper by Rino
{\it et al}~\cite{rino_93}. Therefore we restrict ourselves at this
place to study the cooling rate distribution of the size of the
rings.

In Fig.~\ref{fig15} we show the probability that a particle is member
of a ring with a given size versus the cooling rate. Note that this
distribution is not the same as the probability to find a ring of
size $n$, since the two distributions differ by a weighting factor of
$n$. A discussion of the latter distribution is given in
Ref.~\cite{vollmayr_buns}. From Fig.~\ref{fig15} we recognize that for
all cooling rates investigated rings of size 6 are the most
frequent ones. This fact can be understood by considering the phase
diagram of silica. At zero pressure the crystalline phase that is
obtained when the system is cooled from the liquid phase is
$\beta$-cristobalite~\cite{stoffler_69}, which has only rings of size
6.  (When the temperature is decreased even further one enters the
phase of $\beta$-tridymite and then $\beta$-quarz, which have rings of
size 6 and 8.) It can be expected that the {\it local} structure of
the amorphous network will be similar to the cristalline network next
to the liquid phase. We thus expect that also in the amorphous phase
the rings of size 6 are the most frequent ones and Fig.~\ref{fig15}
shows that this is indeed the case.

From the figure we also recognize that very short and very long rings
occur only seldom and that their frequency diminishes with decreasing
cooling rate. (Note that we also found very few rings (less than 0.5\%)
of size 2 and 9, which are not shown in the figure.) Thus we find that
also the distribution of the size of the rings, a quantity which
characterizes the structure of the network on the intermediate length
scale, depends noticeably on the  cooling rate and that this dependence
shows that the structure becomes more ordered, i.e approaches the local
topology of cristobalite, when the cooling rate is lowered.

The last quantity we investigate with respect to its cooling rate
dependence is the spectrum of the system. This quantity is interesting
for two reasons: First it can be compared with the results of
experiments and thus it provides a further test on how realistic the
potential is and second it is also of general interest to study the
spectrum in order to gain insight of the dynamical behavior of glasses
at low temperatures 
\cite{lj_cool,badro_95,dos,laird_schober,olig_bermej,bembenek95,vollmayr_mode}.

The spectrum was determined by computing the eigenvalues of the
dynamical matrix given by $(m_j m_k)^{1/2}\partial^2 V(\{
{\bf{r}}_i\})/\partial r_{j,\alpha} \partial r_{k,\beta}$, where $j$
and $k$ are particle indices and $\alpha$ and $\beta$ are the cartesian
components $x,y,z$. In Fig.~\ref{fig16} we show the so obtained
spectrum. In order not to crowd the figure too much we present only
three of the cooling rates investigated (see figure caption for
details). The spectrum of amorphous silica has also been measured in
neutron and Raman scattering experiments and and it was found that it
shows several peaks. Galeener and Lucovsky~\cite{galeener76} report
lines at 495 and 1200cm$^{-1}$ and Kucirkov\'a and
Navr\'atil~\cite{kucirkova94} lines at 460, 802 and 1084cm$^{-1}$ for
their Raman scattering experiments and Carpenter and Price find peaks
at 400, 810, 1070 and 1190cm$^{-1}$ in their neutron scattering
experiment~\cite{price_carpenter}.  The location of these peaks are
included in the figure as well (vertical lines).  We see that the
spectrum has two main features.  The first one is a double peak at high
frequencies and the second one is a broad, relatively featureless
mountain at intermediate and low frequencies.  It should also be noted
that there is a gap at small frequencies which is a finite size effect,
since the acoustic modes with very small frequencies have a wavelength
that exceeds the size of the simulation box.

Let us first discuss the double peak at high frequencies. We see that
the effect of a decreasing cooling rate is to increase significantly
the height of the two peaks as well as to decrease the minimum between
the two peaks.  Furthermore we see a small shift of the positions of
the two peaks to higher frequencies when the cooling rate is
decreased.  We recognize that the location of the two peaks reproduces
well the ones of the experiment and, because of the mentioned shift,
the accordance between experiment and simulation becomes even better
with decreasing cooling rate. Note that it is a nontrivial feature of
the model that the spectrum shows at high frequencies the double peak
structure observed in experiments. Jin {\it et al} have, e.g., found in
their simulation of amorphous silica, in which a three body potential
was used, that at high frequencies {\it three} peaks are
present~\cite{jin_93} and Della Valle and Venuti have
shown~\cite{dellavalle_94} that the potential proposed by Tsunekuki
{\it et al}~\cite{tsuneyuki_88} gives two peaks, but that their
location does not match the one of the experiments as well as we find
it here for the BKS potential.  Thus we conclude that with respect to
this property the BKS potential seems to be superior to the other
potentials investigated so far.

The part of the spectrum at intermediate frequencies shows a relatively
weak dependence on the cooling rate. This is not surprising, since most
of the modes associated with these frequencies are relatively extended
and, since the structure of the system at larger distances is not
strongly affected by the cooling rate, these modes are likely not to be
affected by the cooling rate either. A more detailed investigation of
this point will be published elsewhere~\cite{vollmayr_mode}.

The spectrum we find at intermediate frequencies seems to reproduce the
experimental spectrum less well than the high frequency part in that we
do not see any prominent feature in the range 400-500cm$^{-1}$ which is
in disagreement with experiments. This is probably due to the fact that
in this frequency range most of the modes involve the movement of
several particles, thus they extend over a larger region of space.
Since it is much harder to devise potentials that are able to reproduce
correctly the forces also on the intermediate range distances it is not
surprising that the BKS potential does not do well on this point and it
shares this flaw with the other models as
well~\cite{jin_93,dellavalle_94,mitra82,garofalini84}. In passing we
also note that the spectrum as determined from a simulation with the
original BKS potential, i.e. without the truncation of the short range
part, gives essentially the same spectrum~\cite{badro_priv}, thus
showing that the discrepancy between the experiment and our simulation
is not due to this truncation.

The low frequency part of the spectrum of glassy materials has recently
been the focus of interest of several investigations since it was found
that in this frequency range there exists an excess of harmonic
excitation, the nature of which is still a matter of
debate~\cite{buchenau86,boson_peak}. In Fig.~\ref{fig17} we show the
low frequency part of the spectra for three different cooling rates.
Since it is customary in experiments to plot not the density of states
itself, but the density of states divided by frequency squared, we have
done likewise.  Also included in the figure is the data from neutron
scattering experiments by Buchenau {\it et
al}~\cite{buchenau86,buchenau_priv}. Note that these curves contain no
adjustable parameter. We recognize that qualitatively the results of
the experiment and the one of our simulation is quite similar.
Furthermore we see that the agreement between experiment and simulation
improves with decreasing cooling rate. Because of the above mentioned
gap in the density of state it can, however, be expected that even for
a significantly smaller cooling rate the discrepancy between experiment
and simulation will not disappear. For this to happen it is likely that
one has to investigate system sizes that are significantly larger than
the one used here, which is, however, currently too demanding on the
computer resources.

\subsection{Properties of the system at finite temperatures}
\label{sec3_3}

Having presented in subsection A the cooling rate dependence of the
glass transition and in subsection B the cooling rate dependence of
various properties of the glass at zero temperature we use the
remaining of this section to investigate the {\it equilibrium}
properties of the system at finite temperatures and to compare these
with the ones of the glass.  In particular we want to find out at what
temperature which properties of the glass are frozen in, or in other
words, how the fictive temperature depends on the property. Furthermore
we attempt to understand what the microscopic reason is for the
occurence of the density anomaly.

In order to address these questions we saved some of the configurations
of the system during the cooling run with the slowest cooling rate and
analyzed these configurations at selected temperatures. In particular
we investigated configurations at $T=7000$K, the highest
temperature, at $T=4840$K, the location of the local maximum in the
density, and at $T=3220$K, the temperature of the local minimum in the
density between the temperature of the density maximum and zero
temperature (see Fig.~\ref{fig4}). In the previous subsection we have
concluded that for the smallest cooling rate the glass transition
temperature is around 2900K (see Fig.~\ref{fig2}).  Thus we expect that
the results at the three selected finite temperatures are all
equilibrium results, provided that the glass transition temperature
does not depend too strongly on the quantity investigated (remember
that the glass transition temperatures presented in Fig.~\ref{fig2}
are, strictly speaking, only valid for the enthalpy). We also mention
that, since we made ten independent runs at the smallest cooling rate,
we average at each temperature over these ten independent
configurations.

The comparison of the structure at finite and zero temperatures were
done in two ways. One was to compute for the configurations at finite
temperatures the same quantities that we have investigated at zero
temperature, such as the radial distribution function, and to compare
these quantities with the ones obtained for the glass at zero
temperature.  The second way was to take these configurations, to
determine their intrinsic structure~\cite{intrinsic_structure} by
relaxing the enthalpy via a steepest decent procedure, to compute also for
these {\it relaxed} configurations the quantities that we investigated
for the glass and to compare again. Following Stillinger and Weber we
will call in the following the properties of the system (such as, e.g.,
the density) that are obtained from the relaxed configurations
``intrinsic'' properties (e.g. intrinsic density). Note that doing a
steepest descent procedure at $T=7000$K is equivalent to use a
infinitely fast cooling rate. Thus the so obtained result can also be
related to the ones of the previous subsection.

The first quantity we start with is the enthalpy. For the not relaxed
configuration the value of $H(T_b)$ can be read off from
Fig.~\ref{fig1}. The values of the enthalpy for the relaxed
configurations are included in Fig.~\ref{fig7} as horizontal lines. We
see that the higher the temperature is the higher is the value of the
intrinsic enthalpy. At the two highest temperatures, its value is
larger than the values obtained from the quenches with the different
cooling rates. This is consistent with the observation that these two
temperatures are larger than the glass transition temperatures found
for the various cooling rates (see Fig.~\ref{fig2}). For the lowest
temperature, i.e.  $T=3220$K, the value of the intrinsic enthalpy is
about the same as $H_f$ obtained for the cooling rate $\gamma=3.55\cdot
10^{13}$K/s (see Fig.~\ref{fig7}). From Fig.~\ref{fig2} we recognize
that for this cooling rate the glass transition temperature is about
3050K, which is reasonably close to 3220K, the considered temperature
of the system.  Thus we find that the temperature of the glass
transition, as determined in the way described in Sec.~\ref{sec3_1},
gives a reasonable estimate for the temperature at which the system
falls out of equilibrium with respect to the enthalpy as observable.

A similar result is found for the case of the density.  In
Fig.~\ref{fig8} the horizontal lines give the values of the density of
the relaxed configurations at the three temperatures considered. From
this graph we recognize that for the two higher temperatures the
density is larger than the ones obtained from the quenches with the
different cooling rates. Thus this is again in accordance with the
observation that the glass transition temperature of these quenches is
below these two higher temperatures (see Fig.~\ref{fig2}). For the
lowest temperature, i.e. 3220K, the density we find for the relaxed
configurations is comparable to the one we found for a quench with a
cooling rate in the range $4.44\cdot 10^{12}$K/s $\leq \gamma \leq$
$3.55\cdot 10^{13}$K/s, which corresponds to a range of glass
transition temperatures (see Fig.~\ref{fig2}) of 2900K $\leq T_g \leq$
3050K. Thus also in the case of the density the glass temperature is a
good estimate for the temperature at which the system falls out of
equilibrium.

From Fig.~\ref{fig8} we also recognize that, to within the error
bars, the intrinsic density at 7000K and at 4840K is the same.
Furthermore we see that when the temperature is lowered to 3220K, the
intrinsic density changes relatively strong and then remains almost
constant when the temperature is lowered further (as can be recognized
from the fact that $\rho_f$ depends only weakly on $\gamma$, when the
cooling rate is not too large, even though the corresponding glass
transition temperature is still decreasing in this range of $\gamma$).
Thus we conclude that the intrinsic density can be considered to be
essentially constant for temperatures above 4840K, the location of the
anomaly in the density, and below 3220K, the location of the local
minimum of the density and that the intrinsic density changes mainly in
the temperature range between the local maximum and the local minimum
of the density. Note that this temperature dependence of the
{\it intrinsic} density is in stark contrast with the one of the density.
For the latter we find that it is changing at all temperatures and
that it shows a local maximum and a local minimum whereas the
temperature dependence of the former seems to be much simpler. Thus
it seems that the intrinsic structure of the network changes mainly
in the temperature interval between the mentioned maximum and
minimum.

In order to study this effect in more detail we investigate the radial
distribution function $g(r)$. In Fig.~\ref{fig18} we show this quantity
for the Si-O correlation without the relaxation (Fig.~\ref{fig18}a) and
after the relaxation (Fig.~\ref{fig18}b).  Also included are the curves
we obtained from the quench with the smallest cooling rate (curves
labeled with $T=0$K). From panel a) we recognize that $g(r)$ depends
quite strongly on the temperature in that the height of the individual
peaks and minima become more pronounced. This effect is most
prominent for the first nearest neighbor peak (inset of
Fig.~\ref{fig18}a). This change with temperature takes place throughout
the whole temperature range investigated. This is not the case with the
intrinsic $g(r)$ (panel b). We see that in this case the curves
corresponding to $T=7000$K and $T=4840$K are almost the same.  They
differ, however, significantly from the curve at $T=3220$K, which in
turn is very similar to the curve for $T=0$K.  Thus, as in the case of
the density, we come also here to the conclusion that the intrinsic
structure of the network is changing mainly in the temperature range
between the local maximum and the local minimum of the density.
Similar results were found for the {\it intrinsic} radial distribution
functions for Si-Si and O-O~\cite{vollmayr_phd} and thus we will not
show these functions here.

From Fig.~\ref{fig18}a we recognize that the position of the
individual peaks are essentially independent of temperature.
This is not the case for the $g(r)$ for the Si-Si and the O-O pairs,
which are shown in Fig.~\ref{fig19}. We see that for these radial
distribution functions the location of the first peak in
$g_{\mbox{\footnotesize SiSi}}(r)$ moves to larger values when the
temperature is lowered whereas the position of the second nearest
neighbor peak moves to smaller values. For the $g(r)$ for the O-O
correlation the position of the first nearest neighbor peak is
essentially independent of the temperature whereas a careful inspection
of the figure shows that the position of the second nearest neighbor
peak moves to smaller distances when the temperature is lowered from
$T=7000$K to $T=3220$K and then starts to increase to larger values
when the temperature is reduced further.

From the results just presented we thus see that the temperature
dependence of the various pair correlation functions is quite
complicated. The reason for this is presumably the densification of the
network when the temperature decreases from high temperatures to
$T=4840$K and the subsequent opening up of the network when the
temperature is lowered further. 

Equipped with the radial distribution functions we can identify the
nearest neighbors of every particle via the location of the first
minimum in the radial distribution function at the corresponding
temperature. The values for $r_{\mbox{\footnotesize min}}^{\alpha\beta}$
at finite temperatures are given in Tab.~\ref{tab3}. As in the case of
the radial distribution function we find that these distribution
functions show a relatively regular dependence on
temperature~\cite{vollmayr_phd}. An exception, however, seems to be the
distribution function for the O-O pairs. We find that this
distribution function depends only weakly on temperature for $T\leq
3220$K but then changes strongly when the temperature is lowered to
$T=0$K in that it shifts its maximum from $z=8$ to $z=6$ and becomes
peaked much stronger. The reason for this is likely to be the opening
of the network with decreasing temperature.  However, why this change
is so pronounced and why it takes place in the temperature range below
3220K and not as the other quantities in the temperature range 3220K
$\leq T \leq$ 4840K is unclear.

The distribution functions for the intrinsic coordination numbers show
the usual dependence on temperature in that they show only a weak
temperature dependence for $T\leq 3220$K and $T \geq$ 4840K, and a much
stronger dependence in the temperature range 3220K $\leq T \leq$ 4840K.
Thus also for these quantities the relevant changes take place in the
temperature range between the local maximum and the local minimum in
the density.

The changing of the structure of the network can also be studied well
with the help of the angles between the various atoms, which are shown
in Fig.~\ref{fig20}. The distributions of the
intrinsic angles at $T=4840$K and at $T=3220$K are very similar to the
ones at $T=7000$K and $T=0K$, respectively~\cite{vollmayr_phd}. Thus
also in this case the intrinsic structure is essentially independent of
temperature for $T$ higher than 4840K and for temperatures lower than
3220K. Only in the temperature range 3220K $\leq T \leq$ 4840 does the
distribution of the intrinsic angles change significantly. In contrast
to this we see that the distribution of the angles, i.e. without the
relaxation, depends on $T$ for the whole temperature range. In
particular we find, see Fig.~\ref{fig20}a, that the distribution for the
angle O-Si-O is very broad at high temperatures and becomes gradually
narrower when $T$ is decreased. Thus we find that the tetrahedra are
significantly distorted at high temperature, in accordance with the
observation that the first nearest neighbor peak in
$g_{\mbox{\footnotesize SiO}}(r)$ becomes relatively broad at high
temperatures (see Fig.~\ref{fig18}a).

Also the distribution of the angle Si-O-Si changes significantly with
temperature, Fig.~\ref{fig20}b. The position of the large peak that is
present at $T=0$K moves to smaller angles and becomes much broader when
the temperature increases. Since this angle measures the relative
orientation between two neighboring tetrahedra, this observation is in
accordance with the picture of the densifying network, when the
temperature is increased. The same conclusion can be drawn from the
distribution of the O-O-O angle, shown in Fig.~\ref{fig20}c. The
position of the broad secondary peak, corresponding to the angle
between a oxygen atom on one tetrahedron, a bridging oxygen and a
oxygen on a second tetrahedron (O\#1-O\#3-O\#2 in Fig.~\ref{fig12}),
moves to smaller angles with increasing temperature indicating that the
two tetrahedra move closer to each other. At the same time the main
peak, corresponding to the angle of three oxygen atoms on the same
tetrahedron, decreases its height and becomes broader, showing that the
tetrahedra are more distorted at high temperatures.

In Fig.~\ref{fig20}d we show the distribution of the angle Si-Si-Si.
We see that also in this case the height of the main peak decreases and
its width increases with increasing temperature. More remarkable is the
fact that the small peak that we see at low temperatures at
60$^{\circ}$, grows significantly. When we discussed this peak in the
context of Fig.~\ref{fig14}b we mentioned that this peak is probably
related to the presence of very short rings. Since we find now that the
height of this peak increases with increasing temperature we expect
that the frequency of these rings increases significantly with
increasing temperature, which is, as we will show below, actually the
case.

The last structural quantity we investigate is the distribution of
the size of the rings which is shown in Fig.~\ref{fig21}. We see
that also in this case the distribution of the intrinsic size of
the rings, see inset of Fig.~\ref{fig21}, depends significantly 
on the temperature only in the temperature range 3220K $\leq T \leq$
4840K. The distribution function of the ring size without the relaxation
shows, however, a temperature dependence that extends throughout the
whole temperature range investigated. We see that this distribution
becomes significantly broader when the temperature is increased and
that the main change is that the short rings become more frequent.
This is in accordance with our discussion in the context of
Fig.~\ref{fig20}d. We also note that at high temperatures we find some
``rings'' that have a size $n=1$, by which we denote ``rings'' that are
not closed, i.e. which are dangling bonds. These types of rings
disappear when the temperature is less than 4840K, showing that from
an energetic point of view such configurations are unfavorable.

The final quantity we studied was the spectrum of the system, which
is shown in Fig.~\ref{fig22}. The intrinsic spectrum, shown in
Fig.~\ref{fig22}a, shows that the main effect of finite temperature is
to smear out the double peak structure at high frequencies and to
fill up the gap between this double peak structure and the broad
mountain at lower frequencies. The main change in the form of the
spectrum takes again place in the temperature interval 3220K $\leq T
\leq$ 4840K.

The spectrum at finite temperature is quite different from the
intrinsic one, since the dynamical matrix has also negative
eigenvalues. It is customary to plot the distribution of the square
root of these negative eigenvalues on the negative frequency
axis~\cite{bembenek95}. We see that at finite temperatures the double
peak at high frequencies is reduced to a shoulder of the large
mountain at lower frequencies. This is the case even for $T=3220$K,
i.e. the temperature for which we have found that most structural
properties of the system are very similar to the ones at $T=0$K. Thus
we find that this dynamic quantity shows a much stronger temperature
dependence at low temperatures than the structural quantities. The peak
in the distribution at negative frequencies, however, shows a regular
dependence on temperature, thus being more similar to the structural
quantities.

\section{Summary and Conclusions}
\label{sec4}

We have presented the results of a large scale computer simulation in
which we investigated how the properties of silica glass depend on the
cooling rate with which the glass was produced. Experiments in which
such cooling rate dependencies were investigated have focussed, so far,
only on the {\it macroscopic} properties of glasses, such as the
density or the glass transition temperature~\cite{champagnon}. One of
the main goals of our investigation was to study how the {\it
microscopic} properties of the glass are affected by the cooling rate
and see how their cooling rate dependence compares with the one of
macroscopic properties. The second goal of our study was to test
whether the silica potential proposed by van Beest {\it et al.} (BKS),
which so far has only been used to describe cristalline and pressurized
amorphous silica, is also suitable to model vitreous silica produced
via a quench in temperature.

In our work we first focussed on the macroscopic quantities, in order
to see whether the cooling rate dependence of these quantities show a
similar behavior as the ones observed in real experiments. We found
that this is indeed the case, in that, e.g., the dependence of the
glass transition temperature on the cooling rate is in qualitative 
accordance with the one of real materials. If this observed cooling
rate dependence is extrapolated to experimental cooling rates, this
accordance seems also to be correct in a semiquantitative way.

We also observed that, if the cooling rate is sufficiently small, the
density shows an anomalous behavior in that it has a maximum at around
4800K. Such an anomaly is also found in real silica, although at a
significantly smaller temperature (1500K). This shows that with respect
to this phenomenon the BKS potential is able to give at least a
qualitatively correct description of non-crystalline silica.

By investigating the properties of the glass at zero temperature we
find that the enthalpy, the density and the thermal expansion
coefficient depend significantly on the cooling rate. The densities we
find are in agreement with the ones of real silica and an extrapolation
of the thermal expansion coefficient to experimental cooling rates is
also consistent with the experimental values for this quantity. Thus we
find also in this case that the BKS potential is a good model for real
silica glass.

After having made sure that the BKS potential gives a reasonably good
description of the {\it macroscopic} properties of amorphous silica and
that our simulations are able to reproduce the cooling rate dependence
of the glass transition at least in a qualitative way we thus could
move on to investigate how the {\it microscopic} properties of the
glass depend on the cooling rate. We found that the radial distribution
functions showed a pronounced dependence on the cooling rate in that
the individual peaks become significantly more pronounced with
decreasing cooling rate. From this, and the fact that the first sharp
diffraction peak in the structure factor also shows a significant
cooling rate dependence towards becoming more pronounced with
decreasing cooling rate, we conclude that the structure of the system
at short and intermediate distances becomes more ordered.  This
conclusion is also corroborated by our observation that the
distribution of the bond angles becomes more structured and that the
various coordination numbers show the tendency that the basic units in
the network become more ideal, i.e. to become regular tetrahedra. That
also the intermediate range order of the glass increases with
decreasing cooling rate can also be inferred from the observation that in
the distribution of the size of the rings the frequency of rings of
size six increases with decreasing cooling rate, which shows that the
{\it local} structure of the system appoaches the one of
$\beta$-cristobalite.

Also the spectrum of the system, as computed from the eigenvalues of
the dynamical matrix, shows a noticable dependence on the cooling rate
in that the two main peaks at high frequencies become more pronounced
when $\gamma$ is lowered. This shows that the neighborhoods of the
individual atoms shows less variation from atom to atom with decreasing
cooling rate. In addition we find that the location of these two peaks
is very close to the one observed in experiments, demonstrating that the
BKS potential is reliable with respect to this quantity as well.
Furthermore we have shown that also at low frequencies the spectrum is
in fair agreement with experiment. A more extensive investigation of
the spectrum and the eigenmodes of the system will be presented
elsewhere~\cite{vollmayr_mode}.

Finally we investigated how the structure of the glass at finite
temperatures differs from the one at zero temperature in order to find
out how the glass transition is affecting the temperature dependence of
various quantities.  We find that the radial distribution functions
show a smooth dependence on temperature, thus showing that the glass
transition is not accompanied by a dramatic change in this quantity.
This is the case for most other structural quantities considered.
However, if we look at the {\it intrinsic} quantities we note that they
show a much more pronounced temperature dependence. Roughly speaking we
can say that above and below the glass transition the intrinsic
quantities are essentially independent of temperature, that they
change, however, significantly in the vicinity of the glass transition.
This shows that these quantities are likely to be a sensitive indicator
for when the system is undergoing the glass transition in accordance
with the findings of Stillinger and Weber and J\'onsson and
Andersen~\cite{intrinsic_structure,jonsson88}.

From the temperature dependence of the various structural quantities
we gain some understanding on the nature of the density anomaly.  We
find that the network becomes more compact when the temperature is
lowered from high temperatures to 4800K, the temperature at which the
anomaly is observed. This shrinking is a complicated process in which
certain bond distances increase, whereas others decrease and where also
the distribution of the various angles changes significantly with
temperature. When the temperature is decreased even further the density
decreases again which can be understood from a microscopic point of
view by a change in the distribution of the various angles which lead
to an opening up of the network. The nearest neighbor bond distances,
however, do not change significantly in this temperature range showing
that the relative position of the tetrahedra among each other is more
important for the anomaly than the geometry of the tetrahedra.

To conclude we can say that we have shown that, similar to fragile
glass formers~\cite{lj_cool}, also the properties of strong glasses 
show a noticable dependence on the cooling rate with which the glass
was produced. In particular we showed that such dependencies can
affect the microscopic quantities much more than they affect the
macroscopic ones and that it might therefore also be interesting to
investigate in real experiments how microscopic quantities depend
on the cooling rate. In addition we have shown that the two-body
potential proposed by van Beest {\it et al.} for cristalline silica
is also able to give a surprisingly good description of amorphous
silica, thus making it possible to investigate these types of glasses
in a relatively efficient manner.

Acknowledgements: We thank C. A. Angell for valuable discussions, U.
Buchenau for permitting us to reproduce his data before publication and
S. Klaum\"unzer for informing us on experiments on ion bombardment.  K.
V. thanks Schott-Glaswerke Mainz for financial support through the
Schott-Glaswerke-Fond and the DFG, through SFB 262, for financial
support. Part of this work was done on the computer facilities of the
Regionales Rechenzentrum Kaisers\-lautern.

\newpage

\begin{figure}
\caption{Enthalpy $H$ of the system versus $T_b$, the temperature of
the heat bath, for all cooling rates investigated. Main figure:
Enlargement of the glass transition region. The solid and dashed bold
curves are the smallest and largest cooling rates, respectively. Inset:
Full range of temperature.
\protect\label{fig1}}
\vspace*{5mm}
\par

\caption{Glass transition temperature $T_g$ versus the cooling rate.
The solid line is a fit with the functional form given by
Eq.~(\protect{\ref{eq2}}) and the dashed line with a power law.
\protect\label{fig2}}
\vspace*{5mm}
\par
\caption{Specific heat versus $T_b$ for all cooling rates investigated.
The solid and dashed bold curves are the smallest and largest cooling
rates, respectively.
\protect\label{fig3}}
\vspace*{5mm}
\par
\caption{Density of the system versus $T_b$ for all cooling rates
investigated.  The solid and dashed bold curves are the smallest and
largest cooling rates, respectively. Note the presence of a local
maximum in $\rho$ at temperatures around 4500K if $\gamma$ is small.
\protect\label{fig4}}
\vspace*{5mm}
\par
\caption{Thermal expansion coefficient at constant pressure versus
$T_b$ for all cooling rates investigated.  The solid and dashed bold
curves are the smallest and largest cooling rates, respectively.
\protect\label{fig5}}
\vspace*{5mm}
\par
\caption{Thermal expansion coefficient at $T=0$K versus cooling rate.
\protect\label{fig6}}
\vspace*{5mm}
\par
\caption{Enthalpy after the quench versus the cooling rate (open
circles). The solid and dashed curves are fits with the functional
forms given by Eqs.~(\protect{\ref{eq3}}) and (\protect{\ref{eq4}}),
respectively. The three horizontal lines are the value of the enthalpy
of the relaxed configuration at $T_b=7000$K, $T_b=4840$K and
$T_b=3220$K (top to bottom). See text for details.
\protect\label{fig7}}
\vspace*{5mm}
\par
\caption{Density after the quench versus the cooling rate. The three
horizontal lines are the value of the density of the relaxed
configuration at $T_b=7000$K, $T_b=4840$K and $T_b=3220$K (top to
bottom). See text for details.
\protect\label{fig8}}
\vspace*{5mm}
\par

\caption{Radial distribution function. a) $g_{\mbox{
SiSi}}(r)$. Main figure: The slowest (solid curve) and fastest (dashed
curve) cooling rate. The vertical dotted lines give the position of the
peaks as determined from experiments (see Tab.~I).
Inset: Enlargement of the second nearest neighbor
peak for four selected cooling rates. b) $g_{\mbox{
SiO}}(r)$ and $g_{\mbox{OO}}(r)$ for the slowest (solid
curves) and fastest (dashed curve) cooling rate. Inset: Enlargement of
the second and third nearest neighbor peak.
\protect\label{fig9}}
\vspace*{5mm}
\par
\caption{Partial structure factors for the slowest (solid curve) and
fastest (dashed curve) cooling rate. Insets: Enlargement of the
first sharp diffraction peak. a) Si-Si correlation. b) Si-O
correlation. c) O-O correlation. 
\protect\label{fig10}}
\vspace*{5mm}
\par
\caption{Partial coordination numbers versus the cooling rate. a)
Si-O pairs. b) O-Si pairs. c) Si-Si pairs. d) O-O pairs. Note the 
different scales for the various curves.
\protect\label{fig11}}
\vspace*{5mm}
\par
\caption{Schematic representation of two corner-sharing tetrahedra.
\protect\label{fig12}}
\vspace*{5mm}
\par
\caption{Difference between the partial coordination numbers and the
prediction of the factorization approximation. a) Si-Si pairs. b) O-O
pairs.
\protect\label{fig13}}
\vspace*{5mm}
\par
\caption{Distribution function of various angles and cooling rates.
a) Angle O-Si-O for all cooling rates investigated. The bold solid and
dashed curves correspond to the slowest and fastest cooling rate,
respectively. The vertical line is the experimental value from
Refs.~[43,55,56].
b) Angles O-O-O, Si-Si-Si and Si-O-Si for the slowest
(solid curves) and fastest (dashed curves) cooling rates investigated.
The vertical lines are the experimental values from Refs.~[43,54-56].
\protect\label{fig14}}
\vspace*{5mm}
\par
\caption{Probability that a particle is member of a ring of size $n$
versus the cooling rate.
\protect\label{fig15}}
\vspace*{5mm}
\par
\caption{Spectrum of the system for three different cooling rates:
$\gamma=1.14 \cdot 10^{15}$K/s (bold dashed line), $\gamma=7.10 \cdot
10^{13}$K/s and  $\gamma=4.44 \cdot 10^{12}$K/s (bold solid line). The
vertical lines give the location of the peaks as determined in the
experiments of Ref.~[63-65].
\protect\label{fig16}}
\vspace*{5mm}
\par
\caption{Density of states divided by $\nu^2$ for three different
cooling rates:  $\gamma=1.14 \cdot 10^{15}$K/s (bold dashed line),
$\gamma=7.10 \cdot 10^{13}$K/s and  $\gamma=4.44 \cdot 10^{12}$K/s
(bold solid line) Also included is the result from neutron scattering
experiments~[68,70]
\protect\label{fig17}}
\vspace*{5mm}
\par
\caption{Radial distribution function $g(r)$ for the Si-O
correlation at $T=7000$K, $T=4840$K and $T=3220$K. Also included is
the $g(r)$ as obtained from quenching the system to $T=0$K with the
smallest cooling rate. a) Without relaxation. b) With relaxation.
\protect\label{fig18}}
\vspace*{5mm}
\par
\caption{Radial distribution function $g(r)$ for the Si-Si (a) and the
O-O (b) correlation at $T=7000$K, $T=4840$K and $T=3220$K (without
relaxation).  Also included is the $g(r)$ as obtained from quenching
the system to $T=0$K with the smallest cooling rate.
\protect\label{fig19}}
\vspace*{5mm}
\par
\caption{Distribution function for various angles at $T=7000$K,
$T=4840$K and $T=3220$K (without relaxation).  Also included is the
distribution function  as obtained from quenching the system to $T=0$K
with the smallest cooling rate. a) O-Si-O. b) Si-O-Si. c) O-O-O. d)
Si-Si-Si.
\protect\label{fig20}}
\vspace*{5mm}
\par
\caption{Distribution of the size of the rings at $T=7000$K,
$T=4840$K and $T=3220$K.  Also included is the distribution function
as obtained from quenching the system to $T=0$K with the smallest
cooling rate. Main figure: Without relaxation. Inset: With relaxation.
\protect\label{fig21}}
\vspace*{5mm}
\par
\caption{Spectrum of the system at $T=7000$K, $T=4840$K and $T=3220$K.
Also included is the spectrum as obtained from quenching the system to
$T=0$K with the smallest cooling rate. The vertical
lines gives the location of the peaks as determined in the experiment
of Ref.~~[62-64]. 
a) With relaxation.  b) Without relaxation. The distribution for
negative frequencies corresponds to negative eigenvalues of the
dynamical matrix.
\protect\label{fig22}}
\vspace*{5mm}
\par
\end{figure}

\begin{table}
\begin{center}
\begin{tabular}{|cc||c||c|c|}
\hline
\rule[-3mm]{0mm}{2mm}
& & Simulation [\AA]  & \multicolumn{2}{c|}{Experiment [\AA]} \\
\hline  \hline
\rule[-3mm]{0mm}{5mm}
 SiO  & 1. peak & 1.595(5)  & 
1.608 (\cite{grimley_90}) & 1.620 (\cite{mozzi_69}) \\
      & 2. peak & 4.12(1)   & 
                          & 4.15  (\cite{mozzi_69}) \\
 OO   & 1. peak & 2.590(5)  & 
2.626 (\cite{grimley_90}) & 2.65  (\cite{mozzi_69}) \\
     & 2. peak & 5.01(2)    & 
                          & 4.95  (\cite{mozzi_69}) \\
 SiSi & 1. peak & 3.155(10) & 
3.077 (\cite{konnert_73}) & 3.12 (\cite{mozzi_69})  \\
     & 2. peak  & 5.05(5)   & 
                          & 5.18 (\cite{mozzi_69})  \\
\hline
\end{tabular}
\caption{Location of the first and second nearest neighbor peak in the
radial distribution function $g(r)$. The numbers in parenthesis in the
second column give the error in units of the last digit.
\label{tab1}
    }
\end{center}
\end{table}

\begin{table}
\begin{center}
\begin{tabular}{| c || c || c |c |c | c|}
\hline
& Simulation & \multicolumn{4}{c|}{Experiment}    \\
\cline{1-6}
&  $\gamma=4.4\cdot 10^{12}$K/s & Ref.~\cite{mozzi_69} & 
Ref.~\cite{pettifer_88}  & Ref.~\cite{coombs_85} &
Ref.~\cite{daSilva_75} \\
\hline  \hline
\rule[-3mm]{0mm}{8mm}
OSiO  & 108.3$^\circ$ 
(12.8$^\circ$) & 109.5$^\circ$   & & 109.7$^\circ$ & 109.4$^\circ$\\
SiOSi & 152$^\circ$ (35.7$^\circ$) & 144$^\circ$ (38)$^\circ$ & 142$^\circ$ (26$^\circ$)& 
144$^\circ$, 152$^\circ$ & 153$^\circ$ \\
\rule[-3mm]{0mm}{8mm}
& & & & & \\
\hline
\end{tabular}
\caption{Location and, in parenthesis, the full width at half
maximum of the angles O-Si-O and Si-O-Si as determined from the
simulation and experiments.
\label{tab2}
         }
\end{center}
\end{table}

\begin{table}
\begin{center}
\begin{tabular}{|c||c|c|c|}
\hline
\rule[-3mm]{0mm}{2mm}
$T$ & $r_{\mbox{\footnotesize min}}^{\mbox{\footnotesize SiSi}}$ [\AA] & 
$r_{\mbox{\footnotesize min}}^{\mbox{\footnotesize SiO}}$ [\AA] &
$r_{\mbox{\footnotesize min}}^{\mbox{\footnotesize OO}}$ [\AA] \\
\hline  \hline
\rule[-3mm]{0mm}{5mm}
7000K & 3.80 & 2.50 & 3.70 \\
4838K & 3.37 & 2.50 & 3.60 \\
3220K & 3.37 & 2.30 & 3.25 \\
\hline
\end{tabular}
\caption{Location of the first minimum in the radial distribution function 
$g(r)$. 
\label{tab3}
    }
\end{center}
\end{table}

\end{document}